# Resource Allocation and Relay Selection for Collaborative Communications

Saeed Akhavan Astaneh, Saeed Gazor

## Abstract

We investigate the relay selection problem for a decode and forward collaborative network. Users are able to collaborate; decode messages of each other, re-encode and forward along with their own messages. We study the performance obtained from collaboration in terms of 1) increasing the achievable rate, 2) saving the transmit energy and 3) reducing the resource requirement (resource means time-bandwidth). To ensure fairness, we fix the transmit-energy-to-rate ratio among all users. We allocate resource optimally for the collaborative protocol (CP), and compare the result with the non-collaborative protocol (NCP) where users transmits their messages directly. The collaboration gain is a function of the channel gain and available energies and allows us 1) to decide to collaborate or not, 2) to select one relay among the possible relay users, and 3) to determine the involved gain and loss of possible collaboration. A considerable gain can be obtained if the direct source-destination channel gain is significantly smaller than those of alternative involved links. We demonstrate that a rate and energy improvement of up to $\left(1+\sqrt[\eta]{\frac{k}{k+1}}\right)^\eta$ can be obtained, where $\eta$ is the environment path loss exponent and $k$ is the ratio of the rates of involved users. The gain is maximum for low transmit-energy-to-received-noise-ratio (TERN) and in a high TERN environment the NCP is preferred.

## Index Terms

Collaboration, relay selection, resource allocation, rate improvement, energy saving, resource efficiency.

## I. INTRODUCTION

In wireless networks, the main interrelated quantities are achievable rate, consumed transmit energy and efficiency of resource. Many recent results [1]–[5] show that collaboration among

Authors are with the Department of Electrical and Computer Engineering, Queen's University, Kingston, Ontario K7L 3N6, Canada, s.gazor, astaneh@queensu.ca.

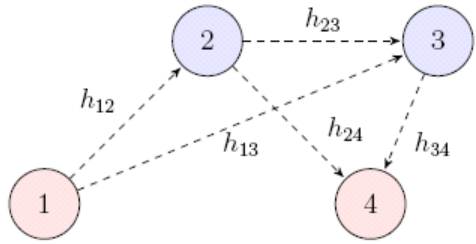

Fig. 1. A collaborative network, the channel energy gain between $i^{\text{th}}$ and $j^{\text{th}}$ user is denoted by $h_{ij}$. Consider three scenarios: 1) the $1^{\text{st}}$ and the $2^{\text{nd}}$ users transmit to the $3^{\text{rd}}$ user, 2) the $1^{\text{st}}$ user transmit to the $3^{\text{rd}}$ user and the $2^{\text{nd}}$ user broadcasts to the $3^{\text{rd}}$ and the 4th users, 3) the $1^{\text{st}}$ to the $3^{\text{rd}}$, the $2^{\text{nd}}$ and the $3^{\text{rd}}$ to the 4th.

users in wireless networks may increase the rate, save on the energy or reduce the resource requirement. However, this is not trivial whether collaboration offers benefit. Here, we ask the question: When collaboration is beneficial?, what are the involved gain or loss from possible collaboration?, and how to selection one relay among the possible candidates? In order to answer the questions, we consider a network of two users (source and relay) intending to send independent information to a destination (See Figure I, the $1^{\text{st}}$ scenario). We propose that the relay user assists the source user only if in a fair way, the collaboration offers benefit in terms of rate, energy or resource. Here, the notion of fairness means that the achievable rates of different users would be proportional to their energy levels. This implies that the ratio of achievable rate over transmit energy for all users are the same. First, we evaluate the effect of collaboration on system performance. Then, we present our relay selection protocol for a general network where we select only one relay user among the possible candidates. In this paper, we extend the results of [6], [7] to the case where rates are not necessarily the same and users are imposed to have fixed ratio of rate over energy.

Most of the existing CPs assume implicitly that a relay is already chosen. In contrary, one might choose only one best relay to assist in the transmission. Several protocols have been proposed to choose the best relay among the potential relay users. Some protocols aim to improve symbol or frame error rate of such a network. Among them, [8], [9] considered symbol error rate of such system where the former studied an amplify-and-forward network and the latter proposed a decode and forward relaying protocol, whereas [10] considered frame error rate of a coded cooperative system. Energy consumption and network lifetime is considered in [11]–[13]. [11], [12] studied decode-and-forward networks and presented several distributed relay selection protocol whereas [13] proposed selective amplify-and-forward relaying protocols. Diversity gain and outage probability have been proposed in relay selection protocols [14]–[18]. While [14]–[17] investigated decode-and-forward and amplify-and-forward network, [18] proposed beamforming to forward data to the destination. However, in some literature the problem of optimal power allocation, relay and relay strategy selection was jointly tackled, using the pricing technique [19], auction theory [20] or convex optimization [21].

In this paper, we study a network of users where all users have independent information to send to corresponding destinations. We first aim to answer the question: What are the involved gains or losses from possible collaboration? In order to answer, we consider a network of three users(See Figure I, the $1^{\text{st}}$ scenario), source, relay and destination, and evaluate the gain of collaboration. We consider a resource allocation problem and study it from three different perspective: 1) rate improvement for a given energy and resource requirement, 2) energy reduction for a given rate and resource requirement, and 3) resource efficiency for a given rate and energy requirement. Then, we are able to answer the following question: Depending on channel gain and transmit energy, when does collaboration offer benefit? We also demonstrate the condition for when users obtain maximum gain from collaboration. We characterize the geometrical conditions under which collaboration is of benefit. Later, we relax the constraint on the number of relay users, and for each case, we present a relay selection protocol. We then move onto a general network topology and examine the proposed protocols in a network where users may wish to communicate with different destinations.

The remainder of the paper is organized as follows. We present the system model and the protocols in Section II. In Section III we study single relay networks and investigate the rate, energy and resource improvement from possible collaboration. We then provide conditions on the location of the relay user for collaboration to be beneficial. In Section IV, we present our relay selection protocols. Extensions to the to the general network with multiple source and relay topology are discussed in Section V. Finally, in Section VI we give our concluding remarks.

## II. System Model and Protocols

Consider the first scenario in Figure I, where we assume that the $1^{\text{st}}$ and $2^{\text{nd}}$ wish to transmit independent messages respectively with rates $R_1$ and $R_2$ to the $3^{\text{rd}}$ user over an additive white Gaussian noise channel (AWGN) and the $2^{\text{nd}}$ user may also assist the $1^{\text{st}}$ user to transmit its messages to the $3^{\text{rd}}$ user. Let denote the energy gain of the communication link between the $i^{\text{th}}$ and $j^{\text{th}}$ user by $h_{ij}$. We assume that the gain of all the channel links are perfectly known to the receivers and transmitters. We also assume that users transmit via a resource division protocol where the $i^{\text{th}}$ user can transmit over a portion $\beta_i$ of available resource.

When the users collaborate, the network is a multi-hopping network where one user receives the messages of another user and forwards the decoded messages to the intended receiver as well as its own messages. Otherwise, they form a multiple access channel, i.e., they transmit directly to the receiver via a resource sharing method.

Following [7], the resource in this paper is defined as the product of used time and the used bandwidth, i.e. $B \times T$. The received energy to noise ratio within the resource slot $\beta_i BT$ can be expressed as $\frac{h_{ij} E_i}{N \beta_i BT}$, where $E_i$ denotes the transmit energy of the $i^{\text{th}}$ user and $N$ denotes the received noise power. Unless otherwise stated, we consider a case where the available resource $BT$ to be unit, i.e. $BT = 1$. Let define the ratio of transmit energy to received noise power (TERN) as $\epsilon_i = \frac{E_i}{N}$. Thus, the achievable rate for is given by

$$R_i = \beta_i \log \left( 1 + \frac{h_{ij} \epsilon_i}{\beta_i} \right). \tag{1}$$

Generally, transmitting at higher energy levels results in higher rates. However we wish to maximize the achievable rates of all users. Similar to [22], [23], we impose the following constraint in order to maintain the fairness,

$$\frac{R_2}{R_1} = \frac{\epsilon_2}{\epsilon_1} \stackrel{\text{def}}{=} k. \tag{2}$$

This constraint ensures fairness among users as the energy spent by users is proportional to their demand for rate. The special case of $k = 1$ is studied in [6], [7].

We consider a half-duplex communication network where each user can either transmit or receive (but not both) at any time and any frequency band. Throughout this paper, we consider two following communication protocols:

- **Non Collaborative protocol** where users transmit directly to the destination via a resource (time and frequency) division method.
- **Collaborative protocol** where over the first resource slot, the $1^{\text{st}}$ user transmits its message and the $2^{\text{nd}}$ user decodes the message of the $1^{\text{st}}$ user. Then, over the $2^{\text{nd}}$ resource slot, the $2^{\text{nd}}$ user re-encodes the decoded message of the $1^{\text{st}}$ user in conjunction with its own message, the $2^{\text{nd}}$ message, and broadcasts the encoded message.

## III. Collaboration in Single Relay Networks

In the following we study some properties of proposed protocols and investigate upper and lower bounds for achievable rates.

### A. Non-Collaborative Protocol (NCP)

In this protocol, during $1^{\text{st}}$ portion of resource slot, i.e. $\beta_1$, the $1^{\text{st}}$ user transmits its message. The receiver, the $3^{\text{rd}}$ user, may be able to decode this message correctly for a maximum rate of $R_1 = \beta_1 \log\left(1 + \frac{h_{13}\epsilon_1}{\beta_1}\right)$. In a similar manner, the maximum rate of the $2^{\text{nd}}$ user which could be decoded reliably at the $3^{\text{rd}}$ user is $R_2 = \beta_2 \log\left(1 + \frac{h_{23}\epsilon_2}{\beta_2}\right)$. Since, we assume that one unit of resource is available, i.e., $\beta_1 + \beta_2 = 1$, hereafter, we denote $\epsilon_1 \stackrel{\text{def}}{=} \epsilon$, $\epsilon_2 = k\epsilon$, $\beta_1 \stackrel{\text{def}}{=} \beta$ and $\beta_2 = 1 - \beta$. Hence, we get the following optimization problem for NCP:

$$\begin{aligned} R_{\text{NCP}} &= \max_{\beta} \left(R_1(\beta) + R_2(1-\beta)\right) \\ \text{s.t.} \quad &\frac{R_2}{R_1} = k \end{aligned} \quad (3)$$

where $R_{\text{NCP}}$ is the achievable sum rate of users and $R_1(\beta) = \beta \log\left(1 + \frac{h_{13}\epsilon}{\beta}\right)$ and $R_2(1-\beta) = (1-\beta) \log\left(1 + \frac{h_{23}k\epsilon}{1-\beta}\right)$. Since $R_1(\beta)$ and $R_2(1-\beta)$ are increasing and decreasing function of $\beta$, respectively, the solution of the above optimization is the unique solution of the following

$$\begin{aligned} R_{\text{NCP}} &= (k+1)\beta \log\left(1 + \frac{h_{13}\epsilon}{\beta}\right) \\ &= \frac{k+1}{k}(1-\beta) \log\left(1 + \frac{h_{23}k\epsilon}{1-\beta}\right) \end{aligned} \quad (4)$$

### B. Collaborative protocol (CP)

In this protocol, over the $1^{\text{st}}$ portion of the resource slot, i.e. $\beta$, the $1^{\text{st}}$ user transmits its messages at rate $R_1$. During this time, The $3^{\text{rd}}$ user is switched off and thus ignores the received signal from the 1$^{\text{st}}$ user. The 2$^{\text{nd}}$ user attempts to decode the messages of the 1$^{\text{st}}$ user. Hence, the maximum achievable rate for the 1$^{\text{st}}$ user is expressed as $R_1 = \beta \log\left(1 + \frac{h_{12}\epsilon}{\beta}\right)$ where, $\epsilon$ denotes the TERN of the first user. Over the remaining portion of resource slot, i.e. $1 - \beta$, the 2$^{\text{nd}}$ user re-encodes the decoded messages of the 1$^{\text{st}}$ user and transmits the messages of the 1$^{\text{st}}$ user as well as its own messages to the intended destination. In fact, during this time, the 2$^{\text{nd}}$ user must transmit at rate of $\frac{k+1}{k}R_2$ to accommodate both data. The maximum achievable rate which may be decoded reliably at the 3$^{\text{rd}}$ user is $R_2 = \frac{k(1-\beta)}{k+1} \log\left(1 + \frac{h_{23}k\epsilon}{1-\beta}\right)$. This yields the following max-min resource allocation problem:

$$\begin{aligned} R_{\text{CP}} &= \max_{\beta} \left(R_1(\beta) + R_2(1-\beta)\right) \\ \text{s.t.} \quad &\frac{R_2}{R_1} = k \end{aligned} \quad (5)$$

where $R_{\text{CP}}$ is the achievable sum rate of users which will be compared with $R_{\text{NCP}}$. In a similar way, the optimal solution is the unique solution of the following equation with respect to $\beta$:

$$R_{\text{CP}} = (k+1)\beta \log\left(1 + \frac{h_{12}\epsilon}{\beta}\right) = (1-\beta)\log\left(1 + \frac{h_{23}k\epsilon}{1-\beta}\right). \quad (6)$$

*C. Rate Improvement for Given Resource and Energy*

In this section, we define the collaboration gain as the ratio of achievable sum rate of the CP to that of the NCP, i.e., $\frac{R_{\text{CP}}}{R_{\text{NCP}}}$. This ratio represents the achievable sum rate improvement of of these protocols. We derive tight upper and lower bounds and study the asymptotic behavior of the collaboration gain at low and high TERN and rate ratio.

Since $R_1(\beta)$ and $R_2(1 - \beta)$ are increasing and decreasing convex and continuous functions of $\beta$, respectively, the maximization (4) is guaranteed to have a unique solution. Unfortunately, this solution has no closed form expression. In Appendix A, we derive the following upper and lower bounds for these achievable rates:

$$R_{\text{NCP}} < \frac{\frac{\log(1+h_{13}(k+1)\epsilon)}{\left(1-\frac{1}{1+h_{13}(k+1)\epsilon}-\log(1+h_{13}(k+1)\epsilon)\right)} + \frac{k\log(1+h_{23}(k+1)\epsilon)}{\left(1-\frac{1}{1+h_{23}(k+1)\epsilon}-\log(1+h_{23}(k+1)\epsilon)\right)}}{\frac{(k+1)}{\left(1-\frac{1}{1+h_{13}(k+1)\epsilon}-\log(1+h_{13}(k+1)\epsilon)\right)} + \frac{k(k+1)}{\left(1-\frac{1}{1+h_{23}(k+1)\epsilon}-\log(1+h_{23}(k+1)\epsilon)\right)}} \quad (7a)$$

$$R_{\text{NCP}} > \frac{\frac{1}{k}\log(1+kh_{23}\epsilon)\log(1+h_{13}\epsilon)}{\frac{1}{k}\log(1+kh_{23}\epsilon) + \log(1+h_{13}\epsilon)} \quad (7b)$$

These bound are tight for high TERN $\epsilon \to \infty$; this is the case where the noise power is negligible compared with the received signal powers. In high TERN regime, the available resource is allocated to the users receive in proportion with their rate demands, i.e., $\lim_{\epsilon \to \infty} \beta = \frac{1}{k+1}$. The lower bound in (7b) is obtained the intersection point of the two lines connecting end points of the rate curves.

Using the same approach, we can find the following bounds for the achievable sum-rate of the CP

$$R_{\text{CP}} < \frac{\frac{\log(1+h_{12}(k+1)\epsilon)}{\left(1-\frac{1}{1+h_{12}(k+1)\epsilon}-\log(1+h_{12}(k+1)\epsilon)\right)} + \frac{(k+1)\log\left(1+h_{23}\frac{k(k+2)}{k+1}\epsilon\right)}{\left(1-\frac{1}{1+h_{23}(k+1)\epsilon}-\log\left(1+h_{23}\frac{k(k+2)}{k+1}\epsilon\right)\right)}}{\frac{k+2}{\left(1-\frac{1}{1+h_{12}(k+1)\epsilon}-\log(1+h_{12}(k+1)\epsilon)\right)} + \frac{(k+1)(k+2)}{\left(1-\frac{1}{1+h_{23}\frac{k(k+2)}{k+1}\epsilon}-\log\left(1+h_{23}\frac{k(k+2)}{k+1}\epsilon\right)\right)}} \quad (8a)$$

$$R_{\text{CP}} > \frac{\frac{1}{k+1}\log\left(1+kh_{23}\epsilon\right)\log\left(1+h_{12}\epsilon\right)}{\frac{1}{k+1}\log\left(1+kh_{23}\epsilon\right)+\log\left(1+h_{12}\epsilon\right)}. \quad (8b)$$

which are tight in the high TERN regime. Since (7a) and (8b), it is easy to see that $\lim_{\epsilon \to \infty} \frac{R_{\text{CP}}}{R_{\text{NCP}}} \geq \frac{k+1}{k+2}$. In addition, from (8a) and (7b), we can see that $\lim_{\epsilon \to \infty} \frac{R_{\text{CP}}}{R_{\text{NCP}}} \leq \frac{k+1}{k+2}$. Thus $\lim_{\epsilon \to \infty} \frac{R_{\text{CP}}}{R_{\text{NCP}}} = \frac{k+1}{k+2}$. Thus the sum rate gain $\frac{k+1}{k+2}$ is smaller than one in the high TERN regime; this means that where large amount of received energy to noise ratio is available the collaborative schemes are not attractive.

In Appendix A, we also derive the following tight bounds for the low TERN regime (small values of $\epsilon$)

$$R_{\text{NCP}} > \epsilon \frac{2h_{23}+2h_{13}-h_{13}^2\epsilon-kh_{23}^2\epsilon-\sqrt{4(h_{23}-h_{13})^2+\epsilon^2\left(h_{13}^2+kh_{23}^2\right)^2+4\epsilon(h_{23}-h_{13})\left(h_{13}^2-kh_{23}^2\right)}}{4}. \quad (9a)$$

$$R_{\text{NCP}} < \min\left\{\log\left(1+h_{13}\epsilon\right), \frac{1}{k}\log\left(1+kh_{23}\epsilon\right)\right\} \leq \epsilon \min\{h_{13}, h_{23}\}. \quad (9b)$$

In addition, the achievable rate is also lower bounded by two end points of the curves, i.e. This upper bound is tight for the low TERN regime, i.e. where the received signal is dominated by noise power. From the above, we conclude that

$$\lim_{\epsilon \to 0^+} \frac{R_{\text{NCP}}}{\epsilon} = \min\{h_{13}, h_{23}\}. \quad (10)$$

Similar to the non-collaborative case, we derive the following upper and lower bounds for CP:

$$R_{\text{CP}} < \min\left\{\log(1+h_{12}\epsilon), \frac{1}{k+1}\log(1+kh_{23}\epsilon)\right\} \leq \epsilon \min\left\{h_{12}, \frac{k}{k+1}h_{23}\right\} \quad (11a)$$

$$R_{\text{CP}} > \epsilon \frac{\frac{2kh_{23}}{k+1}+2h_{12}-h_{12}^2\epsilon-\frac{h_{23}^2k^2\epsilon}{k+1}-\sqrt{4\left(\frac{kh_{23}}{k+1}-h_{12}\right)^2+\epsilon^2\left(h_{12}^2+\left(\frac{kh_{23}}{k+1}\right)^2\right)^2+4\epsilon\left(\frac{kh_{23}}{k+1}-h_{12}\right)\left(h_{12}^2-(\frac{kh_{23}}{k+1})^2\right)}}{4} \quad (11b)$$

July 20, 2018 DRAFT

Thus, we conclude that

$$\lim_{\epsilon \to 0^+} \frac{R_{\text{CP}}}{\epsilon} = \min\{h_{12}, \frac{k}{k+1} h_{23}\}. \tag{12}$$

By combining (10) and (12), we get the following result

$$\lim_{\epsilon \to 0^+} \frac{R_{\text{CP}}}{R_{\text{NCP}}} = \frac{\min\{h_{12}, \frac{k}{k+1} h_{23}\}}{\min\{h_{13}, h_{23}\}}. \tag{13}$$

In addition, It is easy to show that $\frac{R_{\text{CP}}}{R_{\text{NCP}}}$ is always smaller than $\frac{\min\{h_{12}, \frac{k}{k+1} h_{23}\}}{\min\{h_{13}, h_{23}\}}$, i.e. $\frac{R_{\text{CP}}}{R_{\text{CP}}} \leq \frac{\min\{h_{12}, \frac{k}{k+1} h_{23}\}}{\min\{h_{13}, h_{23}\}}$. This means that the rate gain can be greater than unity only if $h_{13} \leq \min\{h_{12}, h_{23} \frac{k}{k+1}\}$. In this case, the maximum rate gain $(\min\{\frac{h_{12}}{h_{13}}, \frac{h_{23}}{\frac{k+1}{k} h_{13}}\})$ is only achievable in low TERN regime.

Now, we examine the collaborative gain when the rate ratio is large. It is easy to see that for large $k$, the optimal $\beta$, which is either the solution of (4) or (6), tends to zero, i.e. $\beta \to 0$. This implies that more resource should be allocated to the higher demanding user. Hence, it is easy to show that $\lim_{k \to \infty} \frac{\log(k)}{k} R_{\text{NCP}} = \lim_{k \to \infty} \frac{\log(k)}{k} R_{\text{CP}} = 1$. Then, it follows that

$$\lim_{k \to \infty} \frac{R_{\text{CP}}}{R_{\text{NCP}}} = 1 \tag{14}$$

On the other hand, if $k$ tends to zero (where the rates of the 1st user is larger than the rate of the 2nd user), the optimal $\beta$ for NCP tends to unity, while for CP tends to zero. Thus, the collaborative gain for small values of $k$, i.e. $k \to 0$, is

$$\lim_{k \to 0^+} \frac{1}{k} \frac{R_{\text{CP}}}{R_{\text{NCP}}} = \frac{h_{23}\epsilon}{\log(1 + h_{13}\epsilon)}. \tag{15}$$

It follows that for small enough rate ratio the achievable rate of NCP is strictly greater than that of CP, i.e, $R_{\text{NCP}} > R_{\text{NCP}}$.

*D. Energy Saving for Given Capacity and Resource*

In the following, we are interested in quantifying the advantage of the collaboration in terms of energy saving. This is in contrast to the previous section where the rate is maximized provided a fixed amount of available energy. Here, we assume that each user require some specified rate $R_i$ and has to allocate TERN proportional to $R_i$. In order to meet these rate requirements, users may collaborate (or not) to use available resource efficiently. Given a unit of shared resource, we minimize the TERN as follows

$$\text{CP} : \begin{cases} \min \epsilon_{\text{CP}}, \\ \text{s.t. } R = \beta \log\left(1 + \frac{h_{12}\epsilon_{\text{CP}}}{\beta}\right) = \frac{1-\beta}{k} \log\left(1 + \frac{h_{23}k\epsilon_{\text{CP}}}{1-\beta}\right) \end{cases} \quad (16a)$$

$$\text{NCP} : \begin{cases} \min \epsilon_{\text{NCP}}, \\ \text{s.t. } R = \beta \log\left(1 + \frac{h_{13}\epsilon_{\text{CP}}}{\beta}\right) = \frac{1-\beta}{k+1} \log\left(1 + \frac{h_{23}k\epsilon_{\text{CP}}}{1-\beta}\right). \end{cases} \quad (16b)$$

Since the rates in (3), (5) are monotonically increasing functions of TERN, thus, it is easy to show that optimization problem (16) is the dual of (3) and (5). This means that under similar channel gains, the TERN collaboration gain (i.e., the ratio of TERN in NCP to that of collaborative one $\frac{\epsilon_{\text{CP}}}{\epsilon_{\text{CP}}}$) obtained from (16) is the same as the rate collaboration gain from (3) and (5). More specifically from this duality, we conclude that

$$\frac{\epsilon_{\text{NCP}}}{\epsilon_{\text{CP}}} \leq \frac{\min\left\{h_{12}, \frac{k}{k+1}h_{23}\right\}}{\min\{h_{13}, h_{23}\}}. \quad (17)$$

Similarly, the maximum gain is obtained when the rate demand is small, i.e., as $R \to 0$.

*E. Resource Efficiency for Given Capacity and Energy*

In the following, we compare the CP and the NCP in terms of the resource usage. We assume that the $1^{\text{st}}$ and $2^{\text{nd}}$ user require rates $R$ and $kR$ under TERN constraints of $\epsilon$ and $k\epsilon$, respectively. The used resource $x$ is the solution of $R = x \log\left(1 + \frac{h\epsilon}{x}\right) \leq h\epsilon$ for a specific rate $R$ and a given amount of energy. Note that we have feasible solution only if $R \leq \epsilon \min\{h_{13}, h_{23}\}$ for the NCP and $R \leq \epsilon \min\{h_{12}, h_{23}\frac{k}{k+1}\}$ for the CP. As the required rates approach these upper bounds the resource usage tends to infinity.

*F. Effect of Network Geometry*

In the following, we investigate the impact of the location of the relay user on the collaboration gain. In particular, we assume that the signal attenuation is governed by geometry of users as $h_{ij} = \frac{1}{d_{ij}^\eta}$ on two dimensional plane, where $d_{ij}$ denotes the distance between the $i^{\text{th}}$ and $j^{\text{th}}$ users. While Cai, Yao and Giannakis [24] examined the achievable minimum energy per bit to investigate the optimal relay placement, here, we focus on collaboration gain and look for the best relay user and protocol which maximizes the collaboration gain, i.e. ratio of achievable rate or transmitted energy or resource, via CP to that of NCP. Our objective is to understand the impact of users relative locations on the collaboration gain. To this end, we investigate the region where transmission via collaboration provides more gain and determine the optimal relay user placement for the proposed protocols. We show that when the relay user is in the vicinity of the source and destination users, collaboration is preferred. We also show that the maximum rate and energy gain of $\left(1 + \sqrt[\eta]{\frac{k}{k+1}}\right)^\eta$ can be obtained.

We assume that in the two dimensional plane, the source, relay and destination are located on $(-\frac{1}{2}, 0)$, $(x, y)$ and $(\frac{1}{2}, 0)$, respectively. Plugging the channel gains as $\frac{1}{d^\eta}$ and $\frac{1}{(1-d)^\eta}$ into the equations (4) and (6), we obtain the rate improvement of both protocols as a function of geometry of relay user. Figure 2 depicts the region where collaboration provide more benefit, i.e. the rate of CP is more than that of the NCP. This figure also depicts the contours of rate gain, where the ratio of achievable rate of protocols is fixed numbers (we plotted for the rate gains of 1, 2 and 4). We observe that as the rate ratio $k$ increases the collaboration contours enlarge. Further increasing the rate ratio, the gain contours reduces. It implies that if the users with middle rate demand have incentive to collaborate with other users.

Since the channel gains are symmetric in two dimensional space, it is clear that the optimal relay user lies on the line connecting the source to the destination. We observe that the gain contours are approximately the intersections of two arcs with the radii $(gc)^{1/\eta}$ and $\left(\frac{k+1}{k}gc\right)^{1/\eta}$ with $gc$ being $gc = \frac{R_{\text{CP}}}{R_{\text{NCP}}}$. In order to find the optimal placement of the relay user we examine the equation (13). It is easy to see that the optimal location is

$$d = \frac{1}{1 + \left(\frac{k}{k+1}\right)^{1/\eta}} \tag{18}$$

where at that point the following maximum rate gain is achievable

$$\frac{R_{\text{CP}}}{R_{\text{NCP}}} \leq \left(1 + \sqrt[\eta]{\frac{k}{k+1}}\right)^\eta. \tag{19}$$

Figure 4 presents the rate improvement from CP and NCP protocols versus the rate ratio of users $k$. We observe that for small rate ratio, the rate improvement is zero and for large values of $k$, the rate improvement tends to unity.

Figure 5 depicts the resource gain of the CP compared with NCP, i.e. $\frac{\beta_{\text{NCP}}}{\beta_{\text{CP}}}$ (20), for a required rate of $0.5h_{13}\epsilon$ versus location of the relay node. We observe that for a given required rate, depending on the relay channel condition, the resource gain is greater than unity. We have noticed that for small rate ratio $k$, CP provides more gain in terms of resource usage. In addition, for small rate ratio, the best location for relay user is almost in the vicinity of the source and destination user.

Figure 6 shows the energy gain of the CP compared with the NCP, i.e. $\frac{\epsilon_{\text{NCP}}}{\epsilon_{\text{CP}}}$ (16), for a given required rate of $R = 0.09 h_{13}$ versus the location of the relay node. Employing the CP, we obtain significant energy savings even for $\eta = 3$, provided that the relay is located appropriately. In contrast to the rate and energy gain, we observe that for higher rate ratio (see Figure 5), users benefit less in terms of resource efficiency. We deduce that only users which are interested in resource efficiency, with less rate requirement, can gain from possible collaboration.

## IV. Collaboration in Multiple Relay Networks

In the following, we propose our relay selection protocols based on the collaboration gain which is introduced in previous section. We use the channel gains to select one relay among the available relay users to participate in collaboration. We note that if the NCP outperforms the collaborative one, we fall back on the NCP, i.e. no relay user would be selected and the source sends its information to the destination directly. Otherwise, the source employs one relay in forwarding its information to the destination. The main objective of the proposed protocols are to achieve higher collaboration gain, higher rate improvement, energy saving or resource efficiency while guaranteeing fairness for all users.

### A. Relay Selection: Rate Improvement and Energy Saving

First, we consider the rate improvement as a criterion to select the best relay. As shown in previous section, the energy minimization problem is dual of the rate maximization problem, hence the relay selection protocol holds for the energy saving as well.

The result in (13) is very intuitive and suggests a strategy in deciding to use collaborate and to choose a relay user among the potential candidates. Given the full CSI, collaboration protocol is preferred if $\epsilon \ll 1$ and $h_{13} \ll \min\{h_{12}, h_{23}\frac{k}{k+1}\}$. In order to maximize the rate gain, the best relay user is the one that maximizes the $\frac{\min\{h_{12}, h_{23}\frac{k}{k+1}\}}{h_{13}}$.

The results in (14) and (15) also provide an attractive guideline that for low and high rate ratio, non CP is preferred. We obtain the collaboration gain for different channel gains. The simulation result shows that for some values of the rate ratio $k$, the collaboration gain is more than unity which for that case, collaboration provides gain.

The equation (18) implies that the best relay user, in order to maximize the rate gain, is located in the vicinity of the source and destination user. We observe that under severe path loss, users benefit more from the proposed collaboration relative to direct transmission. Ochiai, Mitran and Tarokh [5] showed the same result in the context of diversity gain which is not in the scope of this paper. This result also appears very attractive that, in contrast to traditional multi-hopping, appropriately designed collaboration can provide a significant rate gain. Figures 3(a) and 3(b) confirm the above results. This indicates that the best location for the relay user is in the vicinity of the midpoint between the transmitter and the receiver pair. This means that by appropriately selecting the relay user, we efficiently take advantage of the geometrical distribution of users. The optimal location of the relay is almost characterized by (13), which serves for relay selection. Note that by selecting one relay, the multiple relay network becomes a single relay network. Thus, the exact rate improvement or energy saving can be examined as in (6), (4) and (16).

*B. Relay Selection: Resource Efficiency*

Now, we address resource efficiency and the objective is to select a relay user among the potential candidates and to decide wether to collaborate or not. We propose the following procedure:

- Feasibility check: We compare $R$ with $\epsilon \min\{h_{13}, h_{23}\}$ for the NCP and with $\epsilon \min\{h_{12}, h_{23}\frac{k}{k+1}\}$ for the CP. Then, we ignore the protocol which is not feasible.
- Resource usage: If both are feasible, we must choose the protocols with the least resource usage. The resource usages $\beta_{\text{NCP}}$ and $\beta_{\text{CP}}$ are the solutions of

$$\begin{cases} R = \beta_{1,\text{NCP}} \log\left(1 + \frac{h_{13}\epsilon}{\beta_{1,\text{NCP}}}\right) = \frac{\beta_{2,\text{NCP}}}{k} \log\left(1 + \frac{h_{23}\epsilon}{\beta_{2,\text{NCP}}}\right), \\ \beta_{\text{NCP}} = \beta_{1,\text{NCP}} + \beta_{2,\text{NCP}}, \end{cases} \quad (20a)$$

$$\begin{cases} R = \beta_{1,\text{CP}} \log\left(1 + \frac{h_{12}\epsilon}{\beta_{1,\text{CP}}}\right) = \frac{\beta_{2,\text{CP}}}{k+1} \log\left(1 + \frac{h_{23}\epsilon}{\beta_{2,\text{CP}}}\right), \\ \beta_{\text{CP}} = \beta_{1,\text{CP}} + \beta_{2,\text{CP}}. \end{cases} \quad (20b)$$

- Collaborator selection: Similarly, we can use the resource usages for the criterion to select the collaborator among multiple feasible candidates.

## V. Collaboration in General Networks

We can extend the proposed protocols to the multiple relay networks, where more than one user are available to relay the messages of a source toward the destination. As we have shown here, we focus on one relay system and look for the best user to serve as relay to maximize the achievable rate, minimize the energy consumption or utilize the available resource more efficiently. To this end, we provide a rough guideline that if $\epsilon h_{i,j} \gg 1$, often the CP outperforms the NCP. Otherwise, if a fixed rate is required, the feasibility of different scenarios must be verified. Among feasible solutions, we must choose the protocol and relays which provide maximum rate, or maximize savings on resource (20) or on energy (16). For CP, a relay among possible candidates must be selected which maximizes $\min\{h_{23}k/(k+1), h_{12}\} \gg h_{13}$.

For example, suppose that in Figure I the $1^{\text{st}}$ user wishes to send data to the $3^{\text{rd}}$ user, while the $2^{\text{nd}}$ user wishes to broadcast independent messages to the $3^{\text{rd}}$ and $4^{\text{th}}$ users. Using this guideline, the $2^{\text{st}}$ user can collaborate with the $1^{\text{nd}}$ user via acting as relay (the more information, the more incentive to collaborate). In this example the $3^{\text{rd}}$ user has no data to send and thus, ironically, has no incentive to collaborate. So the $2^{\text{nd}}$ user should send his data directly to the 4th user.

So far, we have assumed the same destination for both transmissions. We might relax this constraint easily. For example in Figure I, suppose that the $1^{\text{st}}$ user wishes to send messages to the $3^{\text{rd}}$ user and the $2^{\text{nd}}$ and $3^{\text{rd}}$ users wish to send messages to the 4th user. Using the CP, the $2^{\text{nd}}$ user can act as the relay between the $1^{\text{st}}$ and $3^{\text{rd}}$ users and the $3^{\text{rd}}$ user acts as the relay between the $2^{\text{nd}}$ and 4th users.

We have shown that collaboration have the potential to increase the rate gain of the users by a factor of at most $\left(1 + \sqrt[\eta]{\frac{k}{k+1}}\right)^{\eta}$. This result shows that appropriately choosing the relay user and collaboration protocol considerably save the transmit energy, and also reduce interference amongst the users. This allows more users to transmit simultaneously, which increases the overall network throughput. Our proposed protocols not only improves rate, energy or resource utilization of the involved users, but also have the potential to decrease the overall interference of the network. We have shown that collaboration can mitigate the effects of path loss, thus, users can save transmit energy. This saving reduces interference among users which allows to increase density of users in the network through resource reusing.

## VI. CONCLUSION

We used rate, energy and resource usage as criteria for collaboration and relay user selection. We found the conditions under which the collaboration is preferred for all users. Interestingly, the gain users from collaboration in various terms (increase their achievable rate, reduce their transmit energy or use resources more efficiently) can be more significant at low TERN, where the background noise is strong. Clearly, if the background noise is very weak, the collaboration is less attractive. The relative geometrical location of users (i.e., channel responses) must be considered in the relay selection. Very simple criteria are proposed for relay selection. If the relay is in the vicinity between the source and the destination, collaboration can offer good performance. A maximum rate gain (as well ass energy saving gain) of up to $\left(1 + \sqrt[\eta]{\frac{k}{k+1}}\right)^\eta$ can be obtained provided that a collaboration is established with an appropriately located relay, where $\eta$ is the environment path loss exponent. Furthermore, we present several protocols on how to select the best relay among the possible candidates to maximize the cooperation gain.

## APPENDIX

We refer for a similar proof for the special case of $k = 1$ in [6]. We use the first-order Taylor series approximation at point $\frac{1}{k+1}$ for $R_1(\beta)$ and $R_2(1-\beta)$. The intersection point of the approximate lines gives an upper bound for achievable capacity for the NCP. The coordinates of this intersection point are given by

$$\beta = \frac{1}{k+1} + \frac{\frac{1}{k+1}\log\left(\frac{1+(k+1)h_{23}\epsilon}{1+(k+1)h_{13}\epsilon}\right)}{\log\left((1+(k+1)h_{23}\epsilon)(1+(k+1)h_{13}\epsilon)\right) - \frac{(k+1)h_{13}\epsilon}{1+(k+1)h_{13}\epsilon} - \frac{(k+1)h_{23}\epsilon}{1+(k+1)h_{23}\epsilon}} \qquad (21)$$

and (7a).

To find a lower bound, we can approximate functions in (4) by their second order Taylor series versus $\epsilon$ and obtain $R_{\text{NC}} \geq \max\{h_{13}\epsilon - \frac{h_{13}^2\epsilon^2}{2\beta}, h_{23}\epsilon - \frac{kh_{23}^2\epsilon^2}{2(1-\beta)}\}$. To find a tight bound we solve $(h_{23} - h_{13})\beta^2 + \left(\frac{h_{13}^2\epsilon}{2} + \frac{kh_{23}^2\epsilon}{2} + h_{13} - h_{23}\right)\beta - \frac{h_{13}^2\epsilon}{2} = 0$. This quadratic equation has only one feasible solution in the interval $[0,1]$. This bound is described by (9a) and (22).

$$\beta = \frac{h_{23} - h_{13} - \frac{\epsilon k h_{23}^2}{2} - \frac{\epsilon h_{13}^2}{2} + \sqrt{\left(h_{23} - h_{13} - \frac{\epsilon k h_{23}^2}{2} - \frac{\epsilon h_{13}^2}{2}\right)^2 + 2(h_{23} - h_{13})\epsilon h_{13}^2}}{2(h_{23} - h_{13})} \qquad (22)$$

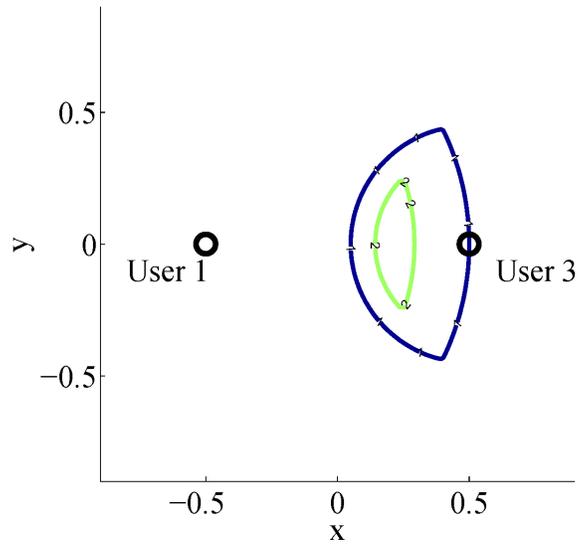

(a)

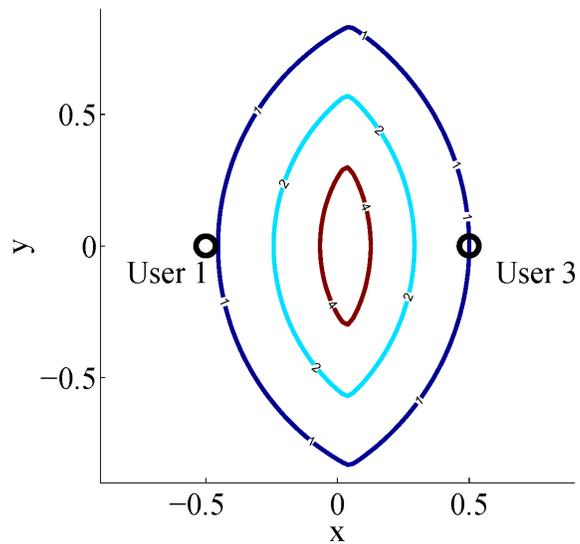

(b)

Fig. 2. Contours of the rate gain $\frac{R_{\text{CP}}}{R_{\text{NCP}}}$ (4), (6) versus relay ($2^{\text{nd}}$ user) location $(x, y)$ for $\epsilon = 0.01$, $h_{ij} = \frac{1}{d_{ij}^{\eta}}$ and $\eta = 3$, (a) $k = 0.1$, (b) $k = 10$.

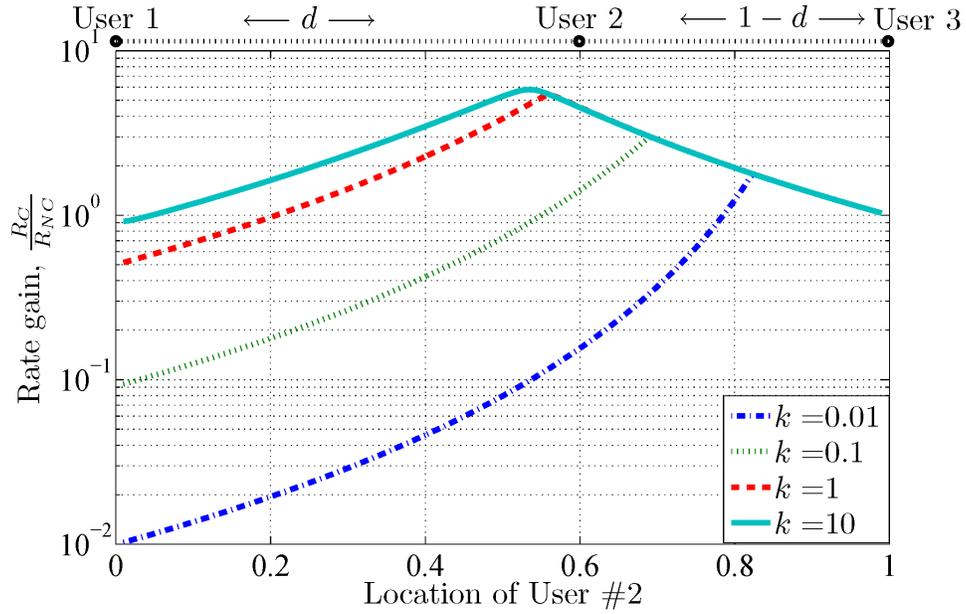

(a)

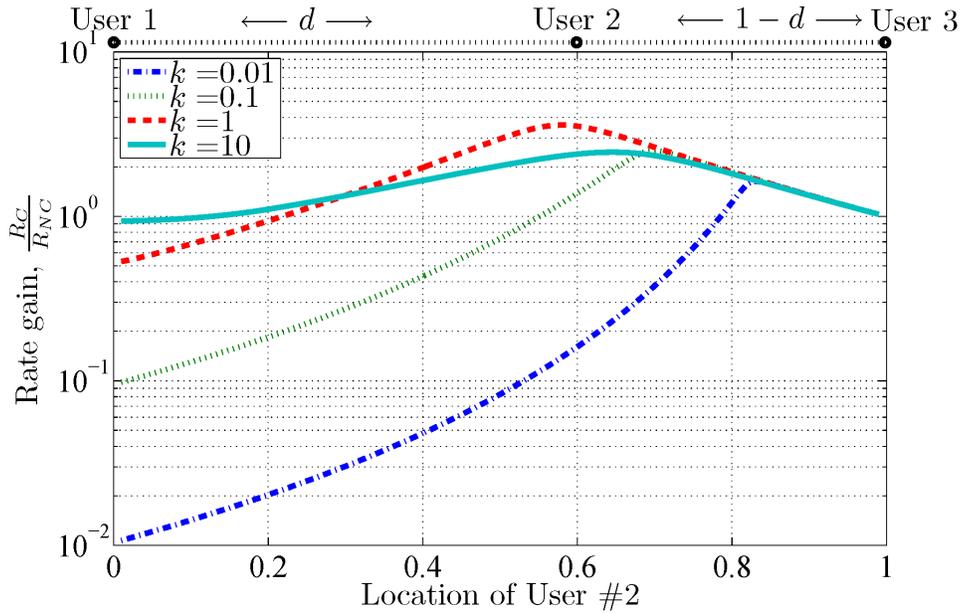

(b)

Fig. 3. Effect of the relay location $d$ on rate improvement $\frac{R_{\text{CP}}}{R_{\text{NCP}}}$ (4), (6) for $h_{12} = \frac{1}{d^\eta}$, $h_{13} = 1$, $h_{23} = \frac{1}{(1-d)^\eta}$, for $\eta = 2$, $k = 0.01, 0.1, 1$ and $10$, respectively, and different TERN values (a) $\epsilon = 0.01$, and (b) $\epsilon = 0.1$.

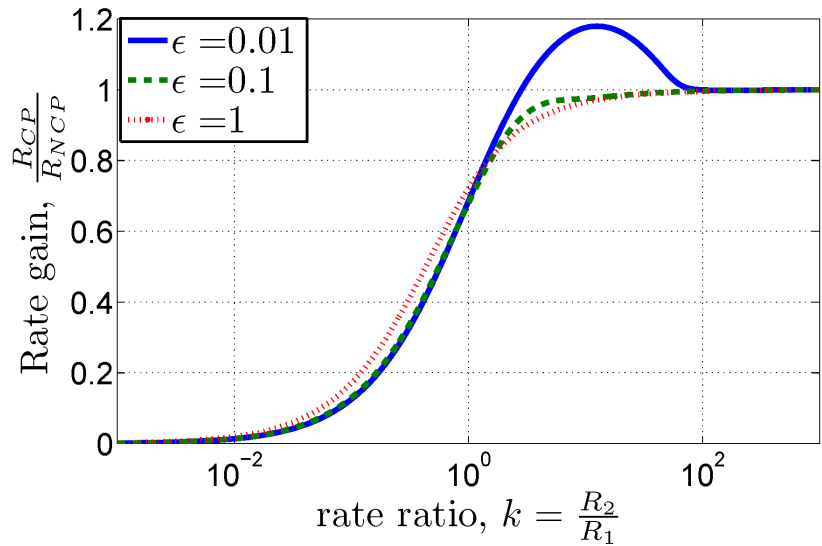

Fig. 4. Effect of rate ratio $k$ on rate improvement, $\frac{R_{\text{CP}}}{R_{\text{NCP}}}$, (4), (6), for $h_{12} = \frac{1}{d^\eta}$, $h_{13} = 1$, $h_{23} = \frac{1}{(1-d)^\eta}$ for a fixed relay location $d = 0.5$ for $\eta = 3$ and and different TERN values $\epsilon = 0.01, 0.1$ and $1$.

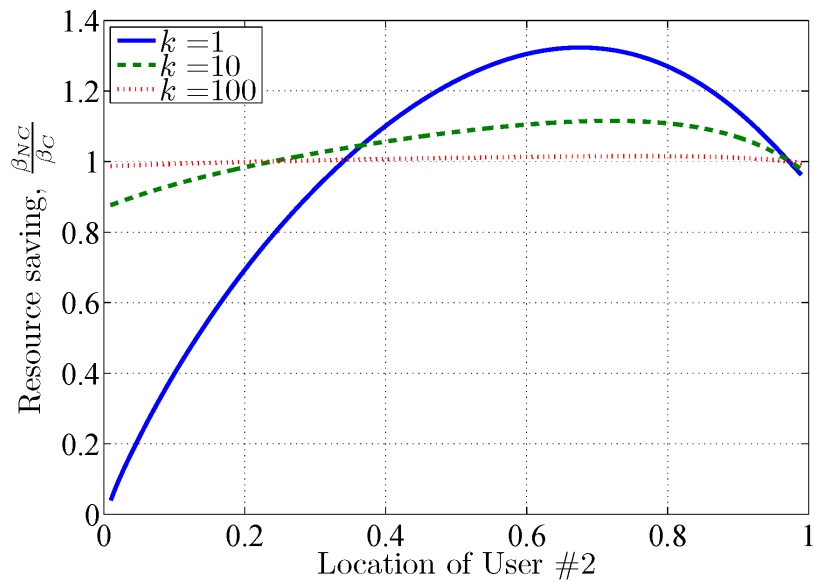

Fig. 5. Ratio of resource usage in CP and NCP $\frac{\beta_{\text{NCP}}}{\beta_{\text{CP}}}$ (20) for $h_{12} = \frac{1}{d^\eta}$, $h_{13} = 1$ and $h_{23} = \frac{1}{(1-d)^\eta}$ versus relay location $d$ for a required rate of $R = 0.5 h_{13}\epsilon$, $\eta = 3$ and $h_{13}\epsilon = 0.01$, and $k = 1, 10$ and $100$.

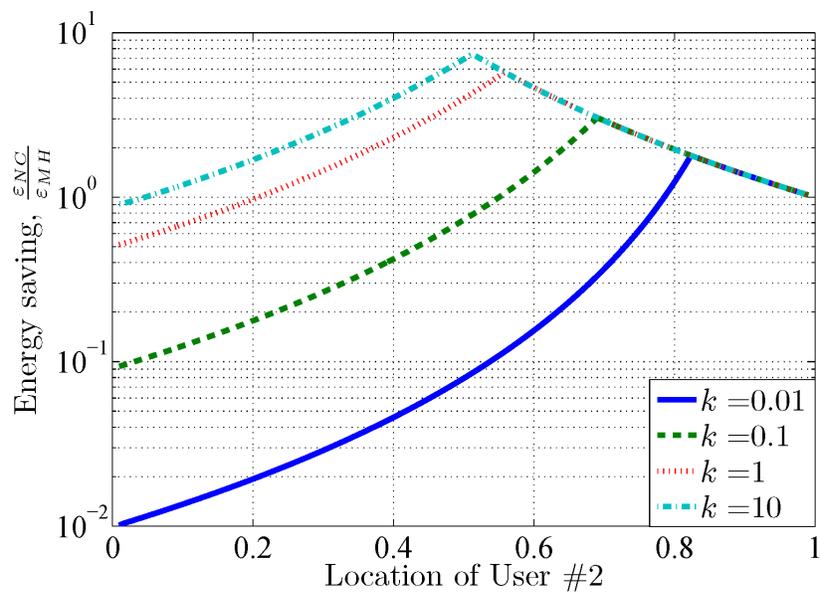

Fig. 6. Ratio of energy usage in CP and NCP $\frac{\epsilon_{\text{NCP}}}{\epsilon\text{CP}}$ (16) for $h_{12} = \frac{1}{d^\eta}$, $h_{13} = 1$, $h_{23} = \frac{1}{(1-d)^\eta}$ and $\eta = 3$ versus relay location $d$ for unit resource and a given required rate of $R = h_{13}/100$, (a) $k = 0.01, 0.1, 1$ and $10$.